\documentclass[10pt]{article}
\usepackage{epsfig,amsmath,amsthm,amsopn,amssymb,doublespace}
\def\d1{{\rho_{1}}}

\begin{document}

\centerline{\bf INSTABILITIES IN A SELF-GRAVITATING MAGNETIZED GAS DISK}
\vskip 3truemm

\centerline{ S. M. Lee and S. S. Hong}

\vskip -1truemm
\centerline{ Department of Astronomy, Seoul National University, Seoul 151-742, 
KOREA}
\vskip 8truemm

\begin{spacing}{1.0}
{\advance\leftskip by 22pt
\advance\rightskip by 22pt
\centerline{\bf ABSTRACT}
\vskip 1truemm

A linear stability analysis has been performed onto a self-gravitating 
magnetized gas disk bounded by external pressure. The resulting dispersion 
relation is fully explained 
by three kinds of instability: a Parker-type instability driven by 
self-gravity, usual Jeans gravitational instability and convection. In 
the direction parallel to the magnetic fields, the magnetic tension
completely suppresses the convection. If the adiabatic index $\gamma$ is 
less than a certain critical value, the perturbations trigger the Parker 
as well as the Jeans instability in the disk.  Consequently, the growth rate 
curve has two maxima: one at small wavenumber due to a combination of 
the Parker and Jeans instabilities, and the other at somewhat larger 
wavenumber mostly due to the Parker instability. In the horizontal direction 
perpendicular to the fields, the convection makes the growth rate increase 
monotonically upto a limiting value as the perturbation wavenumber gets 
large. However, at small wavenumbers, the Jeans instability becomes 
effective and develops a peak in the growth rate curve. Depending on the 
system parameters, the maximum growth rate of the convection may or may 
not be higher than the peak due to the Jeans-Parker instability. Therefore, 
a cooperative action of the Jeans and Parker instabilities can have chances 
to over-ride the convection and may develop large scale structures of 
cylindrical shape in non-linear stage. In thick disks the cylinder is 
expected to align its axis perpendicular to the field, while in thin ones 
parallel to it. \par}
\end{spacing}
\vskip 10truemm

\noindent
{\bf 1. Introduction}
\vskip 2truemm

The Parker instability is one of the most important processes through which 
the Galactic disk may have generated large scale structures. When one suggests 
the instability as a candidate mechanism for making a large scale structure 
in the Galaxy, one should be careful about destructive roles of convection 
(Kim \& Hong 1998). Since growth rate of the convective instability increases 
with decreasing wavelength of perturbation, interstellar medium (ISM) in the 
Galactic disk may get shredded into filamentary pieces by the convection 
before fully developing a structure (Ass\'eo {\it et al.} 1978). In most of 
the previous studies on the Parker instability, externally given gravity was 
taken as a sole source of its driving force. In the present study we instead 
take the self-gravity as the driving force and ignore the external gravity 
from stars. Nagai {\it et al.} (1998) also took the self-gravity into account, 
but they used uniform magnetic fields and considered only isothermal case. 
In making large scale structures the self-gravity ought have played a 
constructive role by triggering the Jeans instability in the medium. 

In this study we model the Galactic ISM as an infinite disk of magnetized 
gas under the influence of its own gravity, and carefully follow up the 
competition among the Jeans, Parker, and convective instabilities. 
We first give adiabatic perturbations to the ISM in an isothermal 
equilibrium, and then perform a linear stability analysis onto the perturbed 
disk to derive the dispersion relation. The $z$-axis is taken 
perpendicular to the disk plane, $y$-axis in the plane along the direction 
of un-perturbed magnetic fields, and $x$-axis perpendicular to the fields. 
All the lengths are normalized to the scale height $H$ of the equilibrium 
disk. The Galactic halo is supposed to bind the ISM disk between 
$z$=$\pm$$z_{\rm a}$, or equivalently $\pm\zeta_{\rm a} (\equiv z_{\rm a}/H)$. 
The normalized wavenumber is denoted by $\nu_x$ and $\nu_y$ for the perturbations 
in the $x$- and $y$-directions, respectively. The growth rate $|\omega|$ is 
normalized by the free-fall time, and we denote the dimensionless rate 
by $|\Omega|$. The system is fully described by the disk thickness, $\zeta_{\rm a}$, 
the ratio of magnetic to gas pressure, $\alpha$, the adiabatic index, 
$\gamma$, boundary conditions, and finally the perturbation wavenumbers, 
$\nu_x$ and $\nu_y$. 

\begin{figure}
\vskip -22truemm
{\centering\epsfig{file=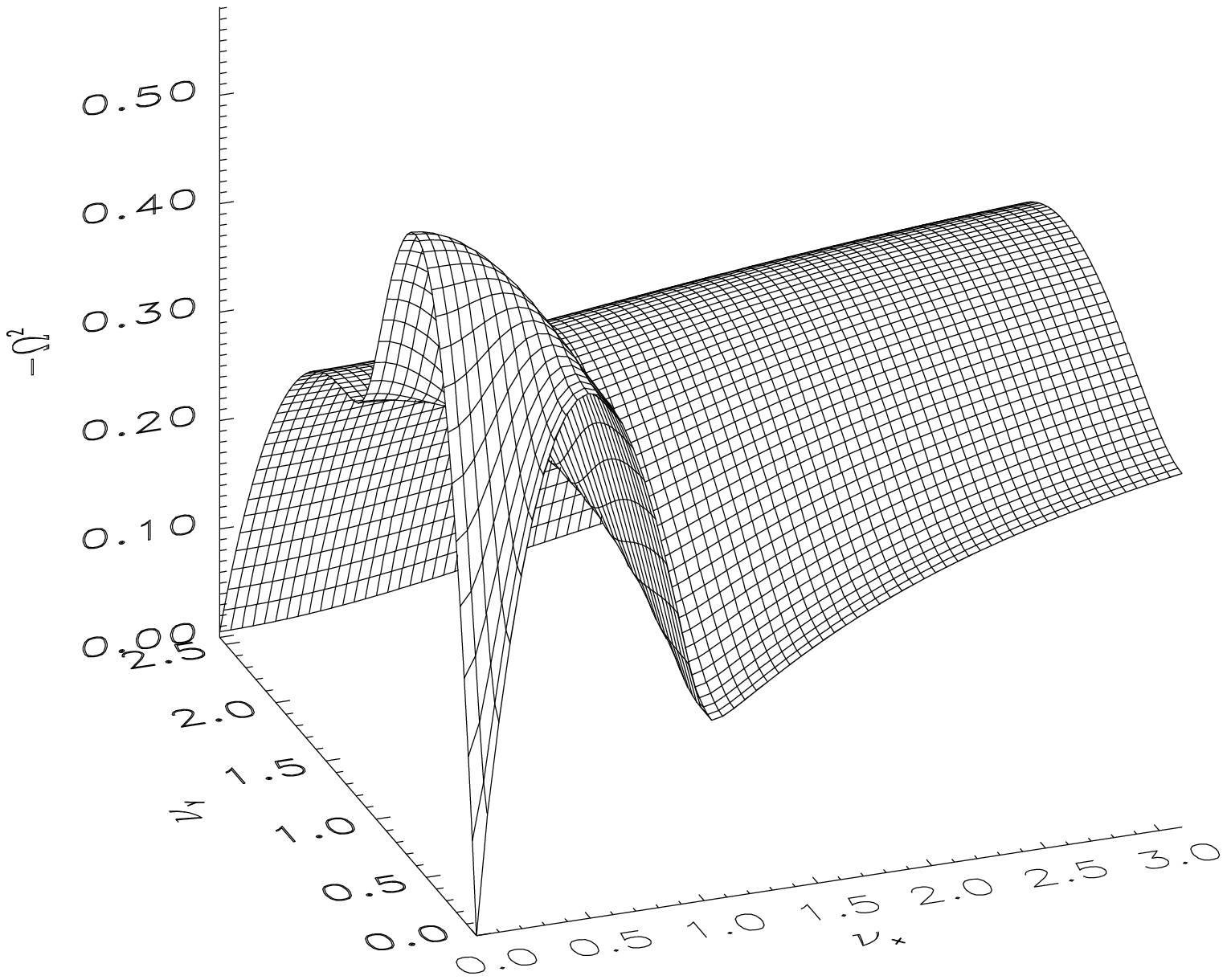,width=12cm,height=9.0cm}}

\vskip -8truemm
\begin{spacing}{0.8}
{\hspace{6truemm}\small {\bf Fig.\enskip 1.--} Dispersion relation for the 
case of a thick disk with $\zeta_{\rm a}$=5.0, $\alpha$=0.1, $\gamma$=0.8. An 
odd symmetry was given to the perturbation at $z$=0. The ordinate is for the 
growth rate, and the two abscissae represent the perturbation wavenumbers.} 
\end{spacing}
\end{figure}
\vskip 4truemm

\noindent
{\bf 2. Dispersion Relations for Thick and Thin Disks}
\vskip 2truemm

Aiming at the gravitational instability, we assigned an odd symmetry to the 
perturbation at $z$=0. Perturbations with even symmetry do not trigger the 
gravitational instability. For the case of a thick disk with $\zeta_{\rm a}$=5.0, 
$\alpha$=0.1, $\gamma$=0.8, and the odd symmetry, we have shown, in 
Figure 1, how the growth rate varies with $\nu_x$ and $\nu_y$. This particular 
set of system parameters is chosen in such a way that we could see all 
the features of the Parker, Jeans and convective instabilities in the 
resulting dispersion relation.

If $\gamma$ $<$ $1+\alpha$, the convection arises in the system. In the 
$x$-direction, the growth rate of the convection increases with increasing 
wavenumber. This is the reason why the ridge height in Figure 1 slowly 
increases, as $\nu_x\rightarrow\infty$, upto the limiting value, $-\Omega^{2}
_{\rm max}=(2/\alpha)\left[1+\alpha+\gamma-2\sqrt{\gamma(1+\alpha)}\right]
\tanh^{2}\zeta_{\rm a}$. However, in the $y$-direction, the convection gets 
completely suppressed by magnetic tension. If $1-\alpha$ $<$ $\gamma$ $<$ 
$1+\alpha$, the magnetic Rayleigh-Taylor instability wouldn't have a chance 
to develop. Since $\gamma$ $<$ $1-\alpha$ in our case, the magnetic 
Rayleigh-Taylor instability can be triggered and yields non-zero growth 
rates all along the $\nu_x$-axis. 

A rather sharp peak in the dispersion curve at ($\nu_x$$\simeq$0.50, 
$\nu_y$=0) is clearly due to the Jeans instability. One can see a similar 
peak at ($\nu_x$=0, $\nu_y$$\simeq$0.61). The latter is  higher than 
the former, because it is due to a combined effect of the Jeans and Parker 
instabilities. The Parker instability driven by the self-gravity has 
brought about the third maximum at around ($\nu_x$=0, $\nu_y$$\simeq$1.4).
 
\begin{figure}
\vskip -22truemm
{\centering\epsfig{file=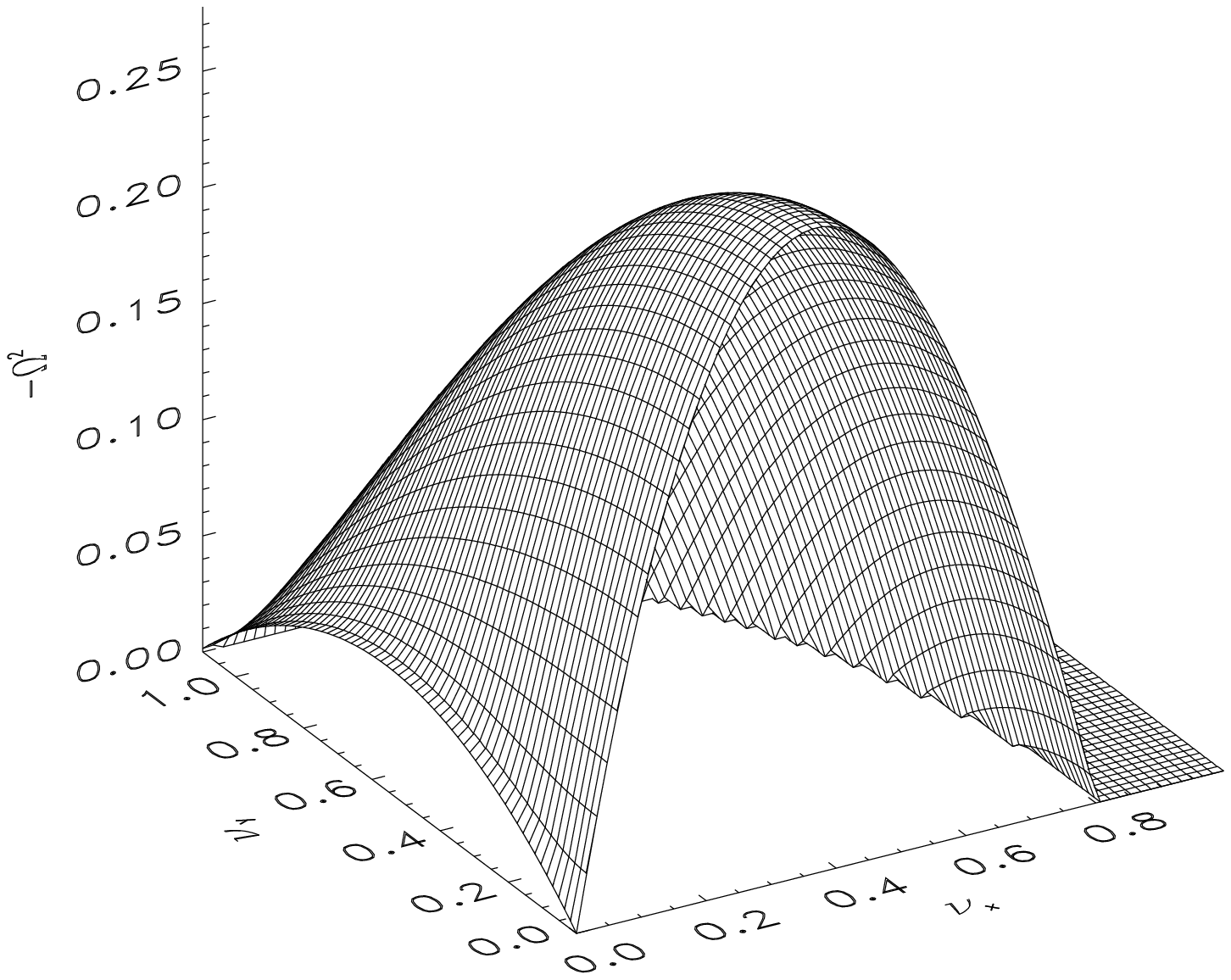,width=12cm,height=9.0cm}}

\vskip -10truemm
{\hspace{6truemm}\small {\bf Fig.\enskip 2.--} The same as in Fig. 1, except 
$\zeta_{\rm a}$=0.1.} 
\end{figure}

Thick disks have enough space for the magnetic fields to bend over so that 
matter can easily slide down. Therefore, the gravity gets an extra boost 
from the fields. This is the reason why the $\nu_y$-axis peak is higher 
than the $\nu_x$-axis one in Figure 1. In thin disks, however, there is 
not enough leeway for the fields to buckle up. Consequently, the fields 
tightly confined in narrow layer hinder, instead of boosting, the system 
not to develop the gravitational instability along the $y$-direction. 
Without being hindered, the system can still develop the Jeans instability 
along the $x$-direction. This makes the $\nu_y$-axis peak lower than the 
$\nu_x$-axis one in Figure 2. Because of the $\tanh^{2}\zeta_{\rm a}$ 
factor, the gravity always over-rides the convection in thin disks. This 
is how Figure 2 becomes so different from Figure 1.

\begin{figure}
\begin{minipage}[b]{.45\linewidth}
\epsfig{file=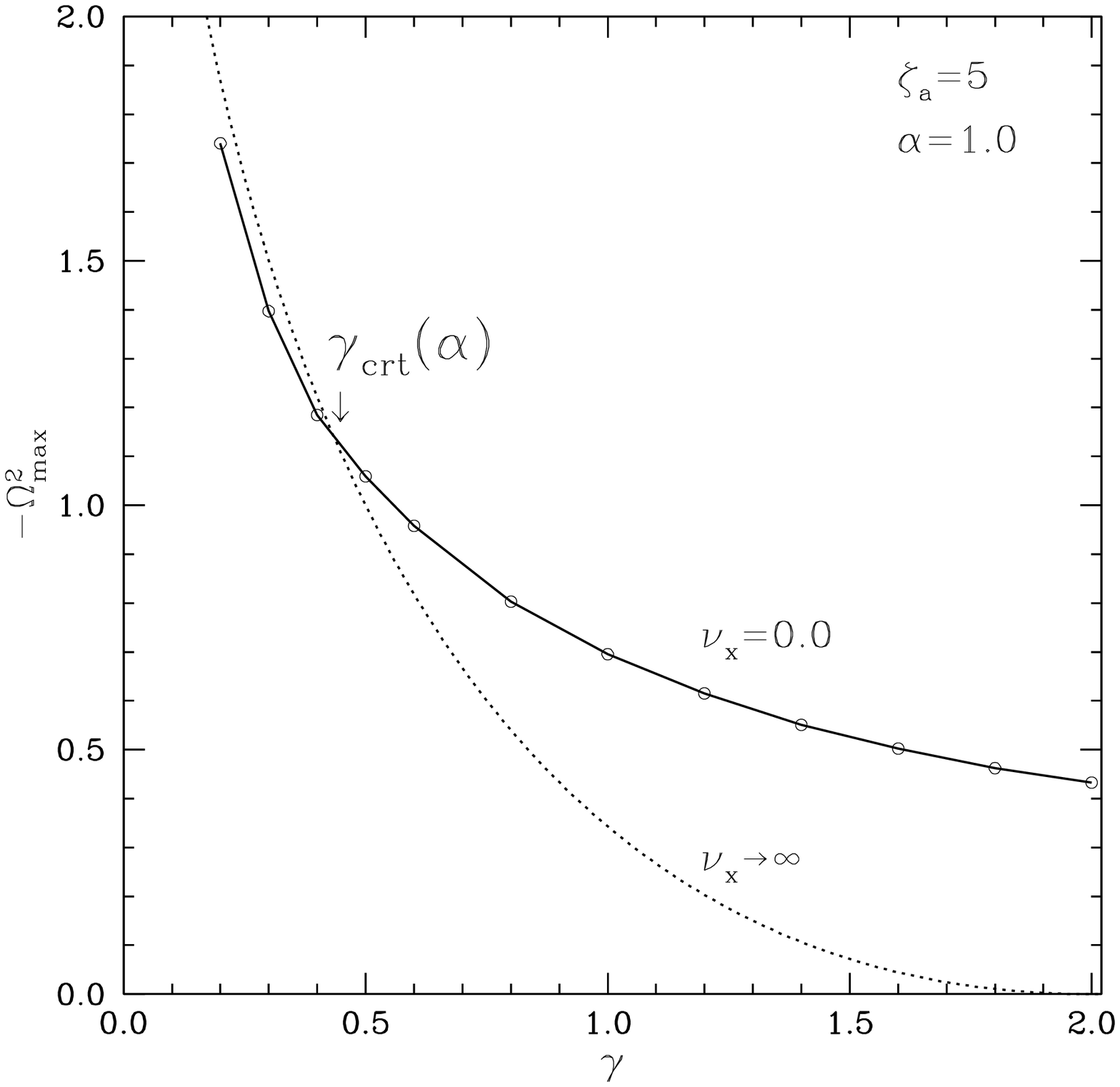, width=\linewidth}
\end{minipage}\hfill
\begin{minipage}[b]{.45\linewidth}
\epsfig{file=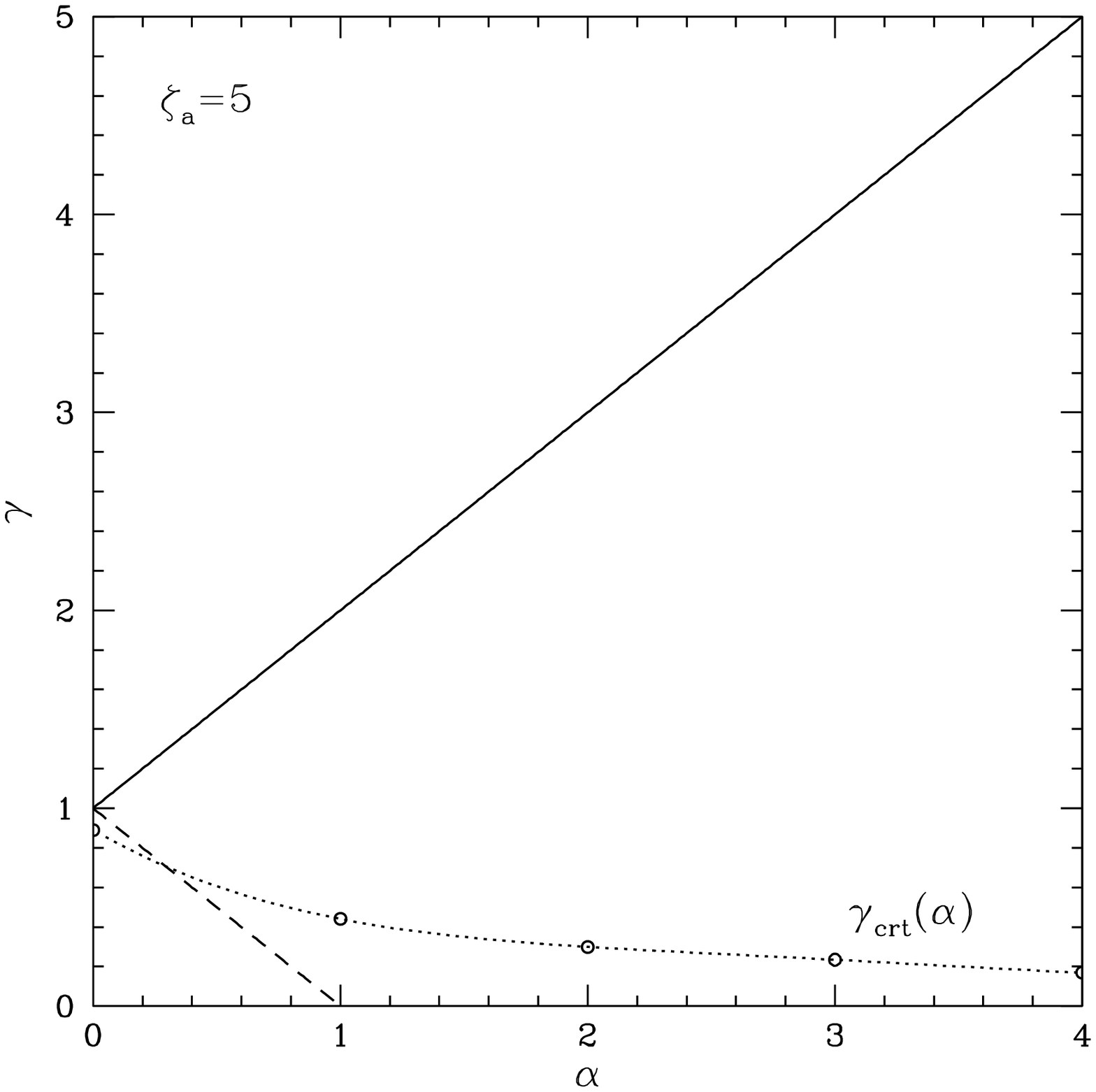, width=\linewidth}
\end{minipage}

\begin{spacing}{0.8}
{\hspace{6truemm}\small {\bf Fig.\enskip 3.--} ({\it left}) Comparison of 
the maximum growth rate between the convection and the Jeans-Parker 
instability. The ordinate and abscissa are for the growth rate and the 
adiabatic index, respectively. When $\gamma$=$\gamma_{\rm crt}(\alpha)$, 
the Jeans-Parker (solid) and the convection (dotted) will grow at the same 
rate. ({\it right}) Instability domains in the ($\alpha$, $\gamma$) plane. 
The solid and dashed lines are for the criteria of convection and magnetic 
Rayleigh-Taylor instability, respectively. The dotted line represents the 
$\alpha$-dependence of the critical adiabatic index.}
\end{spacing}
\vspace{-3mm}
\end{figure}

\vskip 4truemm
\noindent
{\bf 3. Competition between the Jeans-Parker and the Convection}
\vskip 2truemm

In order to see under what conditions the Jeans instability assisted by the 
Parker may win the convection, we have compared their maximum growth rates 
with each other. As can be seen from the left panel of Figure 3, for a given 
$\alpha$, one may find a critical value for $\gamma$, above which the 
Jeans-Parker instability (solid line with open circles) dominates the system 
over the convection (dotted line). Three instability criteria are compared 
with each other in the right panel of the figure: the solid line is for the 
convection, the dashed one for the magnetic Rayleigh-Taylor instability, 
and the dotted line with open circles for the Jeans-Parker instability. In 
the domain below the dashed line both the magnetic Rayleigh-Taylor and 
convection may develop; while in the domain bounded by the dashed and solid 
lines the magnetic Rayleigh-Taylor may not occur ({\it cf.} Newcomb 1961; 
Parker 1967). Above the dotted line with open circles the system forms a 
large scale structure via the Jeans-Parker instability.
\vspace{4mm}

\noindent
{\bf 4. Conclusion}
\vspace{2mm}

The linear stability analysis has marked out in the system parameter space 
those domains where the magnetized gas disk under self-gravity would 
become unstable against the perturbations of large wavelength. If the disk 
is thick, the Parker instability could assist the gravity to generate large 
scale structures of cylindrical shape. Because undular perturbation should 
be applied to the magnetic fields to trigger the Parker instability, the 
structure formed through the Jeans-Parker instability tends to align its 
axis perpendicular to the magnetic fields. In thin disks the undular 
perturbation ought be of short wavelength, and the resulting strong magnetic 
tension wouldn't allow any structures to form perpendicular to the magnetic 
fields. Consequently, if the disk is thin, the self-gravity would drive the 
system to develop large scale structures along the field direction.   
\vskip 1truemm

\noindent
This study was supported in part by a grant from the Korea Research Foundation 
made in the year 1997.

\vspace{5mm}
\noindent
\centerline{\bf REFERENCES}

\noindent
Ass\'eo, E., Cesarsky, C. J., Lachi\`eze-Rey, M., \& Pellat, R. 1978, ApJ, 
225, L21\\
Kim, J., \& Hong, S. S. 1998, ApJ, 507, 254\\
Nagai, T., Inutsuka, S., \& Miyama, S. M. 1998, ApJ, 506, 306\\ 
Newcomb, W. A. 1961, Phys. Fluids, 4, 391\\
Parker, E. N. 1967, ApJ, 149, 535
\vfill\eject
\end{document}